\documentclass[aps,prl,showpacs,twocolumn,superscriptaddress,citeautoscript]{revtex4}
\usepackage{graphics}
\usepackage{graphicx}
\usepackage{amsmath}
\usepackage{amsfonts}
\usepackage{epsfig}
\usepackage{comment,braket}
\usepackage{color}

\def\beq{\begin{equation}}
\def\eeq{\end{equation}}

\begin{document}

\title{Impact of electronic correlations on the equation of state and transport in $\epsilon$-Fe}

\author{L.~V.~Pourovskii}
\affiliation{Centre de Physique Th\'eorique, CNRS, \'Ecole Polytechnique, 91128 Palaiseau, France}
\affiliation{Swedish e-science Research Centre (SeRC), Department of Physics, Chemistry and Biology (IFM), Link\"oping University, Link\"oping, Sweden}
\author{J.~Mravlje}
\affiliation{Josef Stefan Institute, SI-1000, Ljubljana, Slovenija}
\author{M.~Ferrero}
\affiliation{Centre de Physique Th\'eorique, CNRS, \'Ecole Polytechnique, 91128 Palaiseau, France}
\author{O.~Parcollet}
\affiliation{Institut de Physique Th\'eorique (IPhT), CEA, CNRS, URA 2306, F-91191 Gif-sur-Yvette, France}
\author{I.~A.~Abrikosov}
\affiliation{Department of Physics, Chemistry and Biology (IFM), Link\"oping University, Link\"oping, Sweden}

\begin{abstract}
We have obtained the equilibrium volumes, bulk moduli, equations of state of the ferromagnetic cubic $\alpha$ and
paramagnetic hexagonal $\epsilon$ phases of iron
in close agreement with
experiment using an {\it ab
  initio} dynamical mean-field theory approach. 
The local dynamical correlations are shown to be crucial for a
successful description of the ground-state properties of paramagnetic $\epsilon$-Fe.  
Moreover, they enhance the effective mass of the quasiparticles and
reduce their lifetimes across the $\alpha \to \epsilon$ transition leading to a step-wise
increase of the resistivity, as observed in experiment. The calculated
magnitude of the jump is significantly underestimated, which points to non-local correlations.  The implications of
our results for the superconductivity and non-Fermi-liquid behavior of
$\epsilon$-Fe are discussed.
\end{abstract}

\maketitle
Understanding properly pressurized iron is important for the
geophysics of the inner Earth core \cite{Tateno2010} as well as for technological applications of this metal. In the
range 12-16 GPa \cite{Mathon2004,Monza2011} a martensitic transition
from the body-centered-cubic (bcc) ferromagnetic phase ($\alpha$-Fe)
to the hexagonal close-packed (hcp) phase ($\epsilon$-Fe) takes
place. This $\epsilon$-Fe phase, discovered in 1956,
\cite{Bancroft1956} exhibits surprising magnetic and electronic
properties, including superconductivity in the range of pressures from 13 to 31 GPa with
a maximum transition temperature of about 2~K\cite{Shimizu2000} as well as a non-Fermi-liquid normal state observed in the same pressure range \cite{Pedrazzini2006}.

Perhaps the most puzzling aspect of $\epsilon-$Fe is its magnetism, or
rather an unexpected absence of it.  Indeed,
density-functional-theory (DFT) {\it ab initio} calculations within
the generalized-gradient approximation (GGA) predict an
antiferromagnetic ground state for the $\epsilon$ phase, with either
collinear \cite{Steinle-Neumann1999,Steinle-Neumann_2004_2,Friak2008}
or non-collinear \cite{Lizarraga2008} order, which has up to date
eluded experimental detection. While an anomalous Raman splitting
observed in $\epsilon$-Fe \cite{Merkel2000} has been related to a
possible antiferromagnetic order \cite{Steinle-Neumann2004}, no
magnetic splitting has been detected in this phase by M\"ossbauer
spectroscopy down to temperatures of several Kelvins
\cite{Cort1982,Papandrew2006}. A collapse of the ordered magnetism
across the $\alpha-\epsilon$ transition has also been observed in the
x-ray magnetic circular dichroism and in the x-ray absorption
spectroscopy measurements \cite{Mathon2004}.  A spin-fluctuation
paring mechanism that has been proposed for the superconducting phase
\cite{Mazin2002} seems as well incompatible with the large-moment
antiferromagnetism. If the non-magnetic ground state is imposed, the
DFT-GGA total energy calculations predict an equation of state that
drastically disagrees with experiment. The bulk modulus is
overestimated by more than 50\% and the equilibrium volume is
underestimated by 10\% compared to the experimental values
\cite{Steinle-Neumann1999}. Therefore, the ground state properties of
the observed non-magnetic $\epsilon$-Fe remain theoretically
unexplained.

Another puzzling experimental observation is a large enhancement in
the resistivity across the $\alpha$-$\epsilon$ transition, with the
room temperature total resistivity of $\epsilon$-Fe being twice as
large as that of the $\alpha$ phase \cite{Holmes2004}. The
electron-phonon-scattering contribution to resistivity calculated
within GGA is in excellent agreement with the experimental total
resistivity for the $\alpha$ phase \cite{Sha2011}, however, these
calculations predict virtually no change in the resistivity across the
transition to antiferromagnetic hcp-Fe. The enhancement of
resistivity in $\epsilon$-Fe seems likely to be caused by the
ferromagnetic spin-fluctuations as the resistivity $\rho$ follows $\rho \propto
T^{5/3}$  \cite{Jaccard2005,Pedrazzini2006}. This again is at
odds with the antiferromagnetism suggested by the GGA calculations.

All this points to the possible importance of the electronic correlations that
are not correctly incorporated in the local or semi-local DFT. In this paper, we show that
including local dynamical many-body effects significantly improves the
description of iron. Within a local-density-approximation+dynamical mean-field
theory (LDA+DMFT) framework we obtain the ground state properties and
the equation of states (EOS) of both ferromagnetic $\alpha$ and paramagnetic
$\epsilon$ phases of iron as well as the $\alpha-\epsilon$ transition
pressure and volume change in good agreement with experiment. The strength of the electronic
correlations is significantly enhanced at the $\alpha \to \epsilon$
transition. This leads to a reduced binding, which explains the relatively low value of the measured bulk modulus in $\epsilon-$Fe.  The calculated resistivity has a jump at the
transition but the magnitude of the jump is severely underestimated compared with the
experimental value, which points to
additional scattering not present in our local approach.

Previously, the local correlations have been found to improve the
description of the high-temperature paramagnetic $\alpha$-Fe
\cite{Lichtenstein2001,Leonov2011}. Moreover, the recently discovered electronic
topological transition in $\epsilon$-Fe has been successfully
explained by LDA+DMFT but not captured within LDA and GGA
\cite{Glazyrin2013}. However, no attempts to study ground state and transport properties of $\epsilon$-Fe within an LDA+DMFT approach has been reported up to date. 

We have employed a fully self-consistent method \cite{Aichhorn2009,Aichhorn2011} combining a full-potential band structure technique \cite{Wien2k} and LDA for the exchange/correlations with the dynamical
mean-field theory (DMFT) \cite{Georges1996} treatment of the onsite Coulomb repulsion between Fe 3$d$ states. The Wannier orbitals representing Fe 3$d$ states were constructed from the Kohn-Sham states within the energy range from -6.8 to 5.5 eV.
The DMFT quantum impurity problem was solved with the numerically exact hybridization-expansion continuous-time quantum Monte-Carlo (CT-QMC) method \cite{Gull2011} using an implementation based on the TRIQS libraries \cite{TRIQS} package.  The local Coulomb interaction in the spherically-symmetric form was parametrized by the Slater integral $F_0=U=4.3$~eV and Hund's rule coupling $J=1.0$~eV chosen to reproduce the value of the ordered magnetic moment in $\alpha$-Fe of 2.2~$\mu_B$ at its experimental volume. These values are somewhat larger than $F_0=U=3.4$~eV and $J=0.9$~eV obtained by a constrained-random-phase-approximation (cRPA) method in Ref.~\cite{Miyake2009} for $\alpha$-Fe. The 30\% increase with respect to the static cRPA values accounts for the frequency dependence of the Coulomb vertex \cite{Casula2012}. We used the around-mean-field form of the double counting correction~\cite{Czyzyk1994} in this work. We verified that the lowest-energy collinear anti-ferromagnetic order "AFM-II" \cite{Steinle-Neumann1999} collapses at the largest experimental volume of $\epsilon$-Fe in LDA+DMFT-CTQMC calculations with a rotationally-invariant local Coulomb repulsion \cite{Antipov2012}. However, non-density-density terms in the Coulomb vertex increase dramatically the computational cost of CT-QMC and preclude reaching the high accuracy required to extract an equation of state. Hence, we adopted the paramagnetic phase for $\epsilon$-Fe and employed the density-density approximation for the local Coulomb interaction throughout. This allowed us to use the fast "segment picture" algorithm of the CT-QMC \cite{Gull2011} reaching an accuracy of about 0.1 meV/atom in the total energy computed in accordance with Ref.~\cite{Aichhorn2011}.  All our calculations were done for a relatively low temperature $T=290$~K. Thus the phonon and entropic contributions were neglected in the phase stability calculations. 
 The conductivity in DMFT is $\rho^{-1} = 2 \pi e^2 \hbar \int d\omega \sum_k -(\partial f/ \partial \omega) v_k A_k(\omega) v_k A_k(\omega)$ with implicit summation over band indices~\cite{oudovenko_06}. We calculated the band velocities $v_k$ using the Wien2k optics package\cite{Wien2k,Ambrosch2006} and we constructed the spectral functions $A_k(\omega)$ from the
 highly precise  DMFT self-energies (computed using 10$^{11}$ CT-QMC moves) which we analytically continued to the real axis using Pad\'e approximants. 

\begin{table}[ht]
  \caption{Equilibrium atomic volume $V$ (a.u.$^3$/atom) and bulk modulus $B$ (GPa) of bcc and hcp Fe computed by different {\it ab initio} approaches. The $FM$, $PM$, and $NM$ subscripts indicate
   ferromagnetic, paramagnetic, non-magnetic state, respectively, $AFM-II$ is the lowest energy collinear magnetic structure of $\epsilon$-Fe in accordance with GGA calculations of Ref.~\cite{Steinle-Neumann1999}.}.
  \begin{tabular}{lcccc}
    \hline
    {\bf bcc} & LDA+DMFT$_{FM}$ & \multicolumn{2}{c}{GGA$_{FM}$} & expt.$_{FM}$ \\
    \hline
     V &  78.4  & \multicolumn{2}{c}{76.5$^a$, 77.2$^b$, 77.9$^c$} & 78.9 \\
     B &  168  & \multicolumn{2}{c}{187$^a$, 174$^b$, 186$^c$} & 172 \\
    \hline
    {\bf hcp} & LDA+DMFT$_{PM}$ & GGA$_{NM}$ & GGA$_{AFM-II}$ & expt. \\
    \hline
     V &  73.4  & 68.9$^a$, 69$^d$ & 71.2$^d$ & 75.4$^e$ \\
     B &  191  & 288$^a$, 292$^d$ & 209$^d$ & 165$^e$ \\
    \hline
    \multicolumn{5}{l}{
    \begin{minipage}{\columnwidth}%
    \vspace{1mm}
    \begin{flushleft}
    \small LDA+DMFT values are from this work. GGA and exp. values are from a). this work b). \cite{Herper1999}, c). \cite{Friak2008},  d). \cite{Steinle-Neumann1999}, e). \cite{Mao1990} 
    \end{flushleft}
    \end{minipage}%
    }\\
  \end{tabular}
  \label{tabl:B_andV}
\end{table}

\begin{figure*}
\begin{center}
\includegraphics[width=1.9\columnwidth]{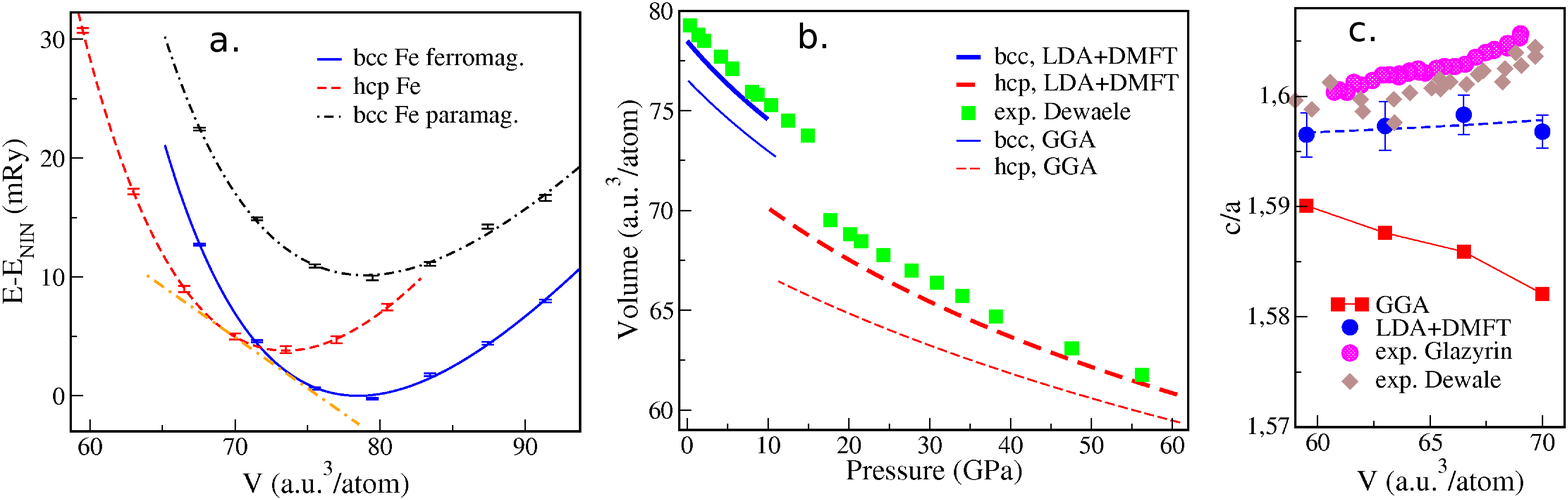}
\end{center}
\caption{\label{fig:eq_st}
(Color online). a). LDA+DMFT total energy vs. volume per atom for bcc (ferromagnetic, solid blue line, and paramagnetic, dot-dashed black line) and hcp (dashed red line) Fe. The error bars are the CT-QMC method stochastic error. The orange long dash-dotted straight line indicates the common tangent construction for the $\alpha-\epsilon$ transition. b). Equations of states (EOS) for ferromagnetic bcc (low pressure) and paramagnetic hcp (high pressure) Fe. Theoretical results are obtained by fitting  the LDA+DMFT (thick line) and GGA (thin line) total energies, respectively,  using the Birch-Murnaghan EOS \cite{Birch1947}. The experimental EOS of iron shown by green filled squares is from Dewaelle et al., Ref.~\cite{Dewaelle2006}. c). The ratio of lattice parameters c/a of $\epsilon$-Fe vs. volume per atom obtained in LDA+DMFT (blue circles, dashed line) and GGA (red squares, solid line). The experimental data are from Dewaelle et al. \cite{Dewaelle2006} (diamonds) and Glazyrin et al.~\cite{Glazyrin2013} (pink circles).
}
\end{figure*}

The obtained LDA+DMFT total energies vs. volume in the $\alpha$ and
$\epsilon$ phases are plotted in Fig~\ref{fig:eq_st}a. Our
calculations predict the ferromagnetic bcc $\alpha$ phase to be the
ground state. The transition to a paramagnetic $\epsilon$-Fe (the
common tangent shown in Fig~\ref{fig:eq_st}a) is predicted to occur at
a pressure $P_c$ of 10~GPa. The paramagnetic $\alpha$ phase is about
10 mRy or 1500~K higher in energy, in good correspondence to the Curie
temperature of $\alpha$-Fe. In Table~\ref{tabl:B_andV} we list the
resulting LDA+DMFT equilibrium atomic volumes and bulk moduli obtained
by the fitting of calculated energy-volume data with Birch-Murnaghan
EOS \cite{Birch1947}. Also shown are GGA lattice parameters and bulk
moduli obtained by us and in previous works, as well as corresponding
experimental values. The LDA+DMFT dramatically improves agreement with
the experiment for paramagnetic $\epsilon$-Fe for both the volume and
bulk modulus, thus correcting the large overbinding error of GGA. The
paramagnetic LDA+DMFT results are still closer to experimental values than the AFM GGA ones.
Hence, even by adopting a magnetic state, which is not observed in experiment, one can only partually 
account for the influence of electronic correlations within the DFT-GGA framework.
For $\alpha$-Fe we obtain a relatively small
correction to GGA, which already reproduces the experimental values
quite well.

In Fig~\ref{fig:eq_st}b we compare the LDA+DMFT and GGA EOS with the one measured from
Ref.~\cite{Dewaelle2006}. Again, for both phases the LDA+DMFT approach
successfully corrects the GGA overbinding error, which is relatively
small in $\alpha$-Fe and very significant in $\epsilon$-Fe. Consequently, LDA+DMFT also reproduces correctly the volume change at the
$\alpha-\epsilon$ transition (about 5\%), which is grossly
overestimated in GGA. The $c/a$ ratio in the hcp $\epsilon$ phase is also affected by the
electronic correlations. As shown in Fig~\ref{fig:eq_st}c, the GGA
calculations predict a reduction of the $c/a$ ratio with increasing
volume, from 1.59 at V=58~a.u.$^3$/atom to 1.58 at
V=70~a.u.$^3$/atom. Within LDA+DMFT the $c/a$ ratio remains almost
constant and is close to the value of 1.60, in good agreement with
experimental measurements \footnote{The peculiarity in the $c/a$ ratio in $\epsilon$-Fe discussed in Ref.~\cite{Glazyrin2013} cannot be resolved in
our calculations. It is an order of magnitude smaller than the error bars due to CT-QMC stochastic error}.

\begin{figure}
\begin{center}
\includegraphics[width=0.95\columnwidth]{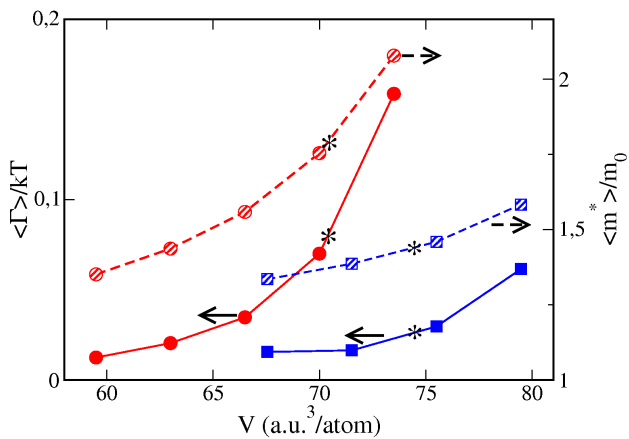}
\end{center}
\caption{\label{fig:Z_Gamma} (Color online) The ratio of the average
  inverse quasiparticle lifetime $\langle \Gamma\rangle$ to
  temperature (the left axis) and the average mass enhancement
  $\langle m^*\rangle/m_0$ (the right axis) vs. volume per atom. The
  solid lines (filled symbols) and dashed lines (hatched symbols) are
  $\langle\Gamma\rangle/T$ and $\langle m^*\rangle/m_0$,
  respectively. The values for bcc and hcp phases are shown by blue
  squares and red circles, respectively. The black stars indicate
  the bcc and hcp atomic volumes at the transition point, respectively.}
\end{figure}

One may see that the ground-state properties (bulk modulus,
equilibrium volume, etc.) of the $\epsilon$ phase are significantly
modified within the LDA+DMFT approach as compared to GGA. In contrast,
for ferromagnetic $\alpha$-Fe those modifications are much weaker. In
order to understand the origin of this difference we have evaluated
the strength of the correlation effects in both phases from the
low-frequency behavior of the local DMFT self-energy $\Sigma(i\omega)$
on the Matsubara grid.  Namely, we computed the average mass
enhancement $\langle m^*\rangle/m_0$ as $\sum_{s}
m^*_{s}N_{s}(E_F)/\sum_{s}N_{s}(E_F)$, where the 
$s$ index designates combined spin and orbital quantum numbers
$\{\sigma,m\}$, $\Sigma_{s}(i\omega)$ and $N_{s}(E_F)$ are
the imaginary-frequency DMFT self-energy and partial density of states
at the Fermi level for orbital $s$, respectively,
$m^*_{s}=1-\left[\left.d\Im\Sigma_{s}(i\omega)/d\omega\right|_{\omega\to
    0}\right]$ is the corresponding orbitally-resolved mass
enhancement. We have also evaluated the average inverse quasiparticle
lifetime $\langle \Gamma\rangle=-\frac{m_0}{\langle m^* \rangle}
\frac{\sum_{s}N_{s}(E_F)*\Im
  \Sigma_{s}(\omega=0)}{\sum_{s}N_{s}(E_F)}$. The
resulting average mass enhancement and inverse quasiparticle lifetime
are plotted in Fig.~\ref{fig:Z_Gamma}. In ferromagnetic $\alpha$-Fe
both quantities slowly decay with decreasing volume and then they
exhibit a large enhancement across the $\alpha-\epsilon$ transition,
indicating a more correlated nature of $\epsilon$-Fe. The latter is
characterized by heavier quasiparticles, a larger
electron-electron scattering and a stronger volume dependence of the
correlation strength as compared to the bcc phase. This analysis
clearly demonstrates that dynamical many-body effects are 
enhanced in the $\epsilon$ phase.

Why are the electronic correlations in $\epsilon$-Fe
 stronger and why does the DFT fail there? The
crucial difference is the magnetism. In $\alpha$-Fe, the
physics is governed by the large static exchange splitting which
easily polarizes the paramagnetic state characterized by a peak in the density of states (DOS)
close to the Fermi energy. Therefore, the spin-polarized DFT-GGA
calculations, which are able to capture this static exchange
splitting, reproduce the ground state properties of ferromagnetic
$\alpha$-Fe rather well. In antiferromagnetic $\epsilon$-Fe obtained within DFT-GGA the static exchange splitting also reduces the bonding 
leading to an improuved agreement with the experimental equation of state.
In contrast, dynamical many-body effects must
be included to describe the experimental paramagnetic state of $\epsilon$-Fe properly. In this
respect we note that the many-body corrections to the total energy and spectral properties 
were shown to be important~\cite{Leonov2011,Katanin2010,Pourovskii2013} for the
high-temperature nonmagnetic state of $\alpha$-Fe as well. If one
suppresses magnetism, $\alpha$-Fe is actually even more correlated than
$\epsilon-Fe$, which is a consequence of its large DOS close to the
Fermi energy\cite{Maglic1973}. This larger DOS implies slower quasiparticles which are
influenced by the interactions, especially the Hund's rule coupling~\cite{georges13}.

\begin{figure}
\begin{center}
\includegraphics[width=0.95\columnwidth]{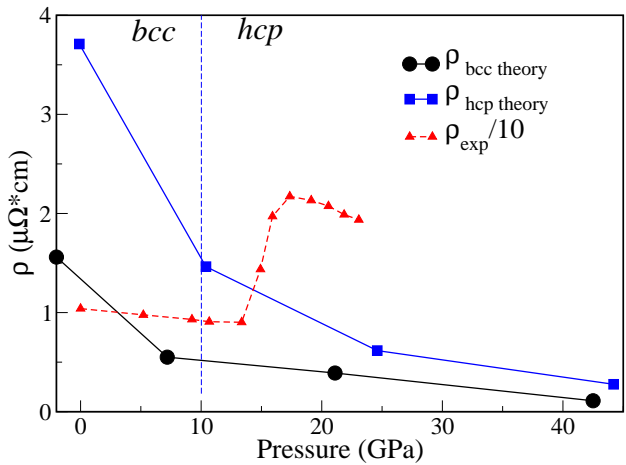}
\end{center}
\caption{\label{fig:resist} (Color online) The electron-electron
  contribution to the resistivity in bcc $\alpha$-Fe (black circles,
  ) and hcp $\epsilon$-Fe (blue squares) computed within LDA+DMFT for T=294~K.  The
  experimental room-temperature resistivity \cite{Holmes2004} (red triangles, dashed
  line) is shown divided by 10 . The vertical dot-dashed line
  indicate the theoretical transition pressure.}
\end{figure}

We now turn to transport. The drop in the quasi-particle lifetime
across the $\alpha-\epsilon$ transition affects the
electron-electron-scattering (EES) contribution to the resistivity
$\rho_\mathrm{el.-el.}$.  We have calculated the evolution of
the room-temperature $\rho_\mathrm{el.-el.}$ versus pressure in both
phases, as displayed in Fig.~\ref{fig:resist}. One may see
that the behavior of $\rho_\mathrm{el.-el.}$ under pressure reflects
that of the inverse quasi-particle lifetime $\Gamma$. The resistivity decays with pressure 
except at the transition point where  a step-wise increase is found. A rapid
enhancement of $\rho_\mathrm{el.-el.}$ in the $\epsilon$ phase at
pressures below $P_c=$10~GPa is in agreement with very recent
measurements \cite{Yadav2013}, in which a large hysteresis in the
$\alpha-\epsilon$ transition was obtained with the transition shifted
to 7~GPa at the depressurization, and a very similar rapid increase in
the total resistivity upon the decrease of pressure was observed in $\epsilon$-Fe for pressures
below the usual experimental $P_c$ of 12-15~GPa.  

Now we discuss a very interesting point: despite the good qualitative
agreement, the magnitude of the resistivity enhancement through the transition
to $\epsilon-$Fe in our calculations underestimates the values found
in the experiment~\cite{Holmes2004}, by a factor of 10, see
Fig.~\ref{fig:resist} \footnote{Note that plot 2 suggests quite long
  scattering lifetimes. The transport scattering time
  $\tau_\mathrm{tr}$ is an average of $-1/[2\Im
  \Sigma(\omega)]$=
  over the window opened by the Fermi function. For $\epsilon$-Fe at
  $T=$295~K one obtains $\tau_\mathrm{tr}\approx 80$fs.}.  The calculations of the
electron-phonon-scattering (EPS) contribution to resistivity in
Ref.~\cite{Sha2011} reproduce the resistivity of the bcc-Fe well but
exhibit almost no change across the transition.  Therefore, the
corresponding jump of the measured total resistivity has to be
attributed to electron-electron scattering. If the experimental
issues, such as the sample thinning can be excluded, then the missing
scattering that we find in comparison with experiments has to be
associated with non-local long-range effects, which are not dealt with in
our calculations.  Interestingly, except for the magnitude, the
pressure dependence  of the resistivity is accounted well by our results. This might
be understood by recognizing that the local correlations that suppress
the coherence scale (the kinetic energy) also make the electronic
degrees of freedom more prone to the effects of the coupling to the
long-range spin-fluctuations.

It is interesting to compare $\epsilon$-Fe and Sr$_2$RuO$_4$, a widely
investigated unconventional superconductor with similar transition
temperature~\cite{mackenzie_rmp_2003}, to make a further
link with spin fluctuations.  Both materials display low-temperature
unconventional superconductivity~\cite{mackenzie_rmp_2003} and several
mechanisms have been discussed to be at its origin~\cite{maeno12}.
In Sr$_2$RuO$_4$ superconductivity emerges from a well-established Fermi liquid with
 $T_\mathrm{FL}=25$K. Local approaches, like the one used in this work, yield
much shorter lifetimes~\cite{Mravlje2011} and a resistivity which agrees with
experiments at low temperatures within 30\%.~\cite{Mravlje_unpub}
The picture is different for $\epsilon$-Fe, which displays a non-Fermi-liquid $T^{5/3}$
temperature dependence of its low-temperature resistivity,~\cite{Jaccard2005}
extending up to a temperature $T^*$ which reaches the peak
$T^*_\mathrm{max}\approx 35$K at a pressure where superconductivity reaches its
maximum \cite{Yadav2013}.
Spin fluctuations, which are believed to be responsible for this
behavior~\cite{Mazin2002} have, hence, a very strong effect in $\epsilon$-Fe. Because
their non-local nature cannot be captured within our framework, we believe that
they are at the origin of the discrepancy between the experimental and our
calculated resistivities.

In summary, including local correlations crucially improves the
theoretical picture of $\epsilon-$Fe by correctly accounting for a set
of experimental observations within the paramagnetic state. This
solves the long-standing controversy between theory and
experiment for this material. The successful description of Fe within the
paramagnetic state, together with an underestimation of the
resistivity found in our local approach, hints at the
importance of spin fluctuations, which supports scenarios relating them to the
origin of superconductivity.

Acknowledgment:
We are grateful  to A. Georges and D. Jaccard for useful discussions. 
The input of X. Deng in the development of the
transport code is gratefully acknowledged.
L.~P.  acknowledges the travel support provided by PHD DALEN Project 26228RM.
J.~M. acknowledges the support of College de France where a part
of this work was done and  the Slovenian research agency program P1-0044.  O.~P. acknowledges support by ERC under grant 278472 - MottMetals.
The
computations were performed on resources
provided by the Swedish National Infrastructure for Computing (SNIC)
at the National Supercomputer
Centre(NSC) and PDC Center for High Performance
Computing.


\end{document}